\documentclass[prb,10pt,twocolumn]{revtex4-1}
\usepackage{amsmath}  
\usepackage{amsfonts} 
\usepackage{graphicx} 

\usepackage{epsfig,amssymb}
\usepackage[colorlinks=true, pdfstartview=FitV, linkcolor=blue, citecolor=blue, urlcolor=blue]{hyperref}

\usepackage{listings}
\usepackage{dcolumn}
\usepackage{bm}

\newcommand{\bea}{\begin{eqnarray}}
\newcommand{\eea}{\end{eqnarray}}
\newcommand{\be}{\begin{equation}}
\newcommand{\ee}{\end{equation}}

\begin{document}

\title{Notes on pulse-width modulation appropriate for a sophomore-level electronics or programming class}

\author{Nathan T. Moore}
\email{nmoore@winona.edu}
\affiliation{Winona State University}

\date{\today}
  
\begin{abstract}
Pulse-width modulation is a nice way to create pseudo analog outputs from a microcontroller (eg, an Arduino), which is limited to digital output signals.  The quality of the analog output signal depends on the filter one uses to average out the digital outputs, and this filter can be analyzed with sophomore-level knowledge of  resistor-capacitor circuits in charging and discharging state.  These notes were used as a text supplement in a sophomore/junior level Microcontroller programming class for second and third-year Physics majors.  I don't know where this material is normally treated in the Physics curriculum and figured others might find the resource useful.  Comments welcome!
\end{abstract}

\maketitle


\section{Background}
Most of the world runs on analog measurement levels.  The amount of coffee in your cup doesn't resolve to an integer number of milliliters, and the sine wave in a residential electric line certainly doesn't have an integer-defined voltage vs time graph.  

What does ``analog" mean?  One way of thinking about the term is to divide the quantity you're measuring, eg, the volume of coffee in your cup, $V_c$, by some smallest measuring unit, eg $dV=0.5mL$ in a graduated cylinder.  If the division always resolves to an integer, the measurement is integer, and if there's often (always) a remainder, the measurement is analog.

\begin{figure}[h]
\begin{center}
\includegraphics[height=5cm]{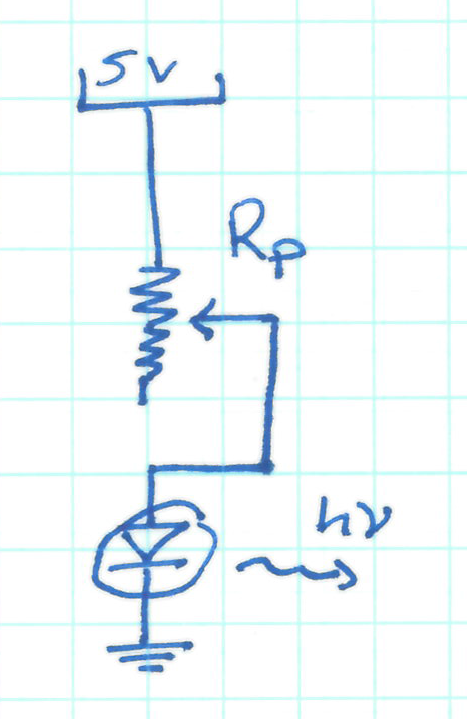}
\label{fig:PWM-POT}
\caption{A simple and energy-inefficient way to dim an LED is to use a potentiometer in series with the LED. Roughly half of the power is thrown away as waste heat.}    
\end{center}
\end{figure}

The microcontroller we've been using in class, an Arduino Uno, \cite{arduino}, only has digital outputs.  Pin 13 can be on, $5v$, or off, $0v$, and the intermediate state, $2.5v$, is impossible.  If you want to dim an LED's brightness, one way to do it is to wire the LED in series with a potentiometer, which mechanically maps electrical resistance to the rotational position of a control.  As you use this ``dial a resistor,'' you mechanically change the (analog) resistance that's in series with the LED.  A higher resistance means less current flows through the series circuit and thus fewer electrons flow through the PN junction in the LED, which means fewer photons are emitted from the LED, which means fewer photons hit your retina, which your brain interprets as ``dimmer."

There are drawbacks to this approach: the potentiometer requires human input, and the series resistor also wastes energy as heat. One workaround to this LED dimming problem we've already used in lab is to toggle the LED anode from on to off at a frequency that's faster than a human eye-brain interface can keep up with.  In practice, if the LED pin is run in an on-off-on-off cycle of about $1ms$ per on or off period, most people see a $50\%$ dimmed LED rather than a blinking LED.  \textbf{Safety note:} recall that varying the on/off period from 1ms to about 20ms can induce seizures in some people - be careful! 

\begin{center}
\begin{figure}
\includegraphics[width=\columnwidth]{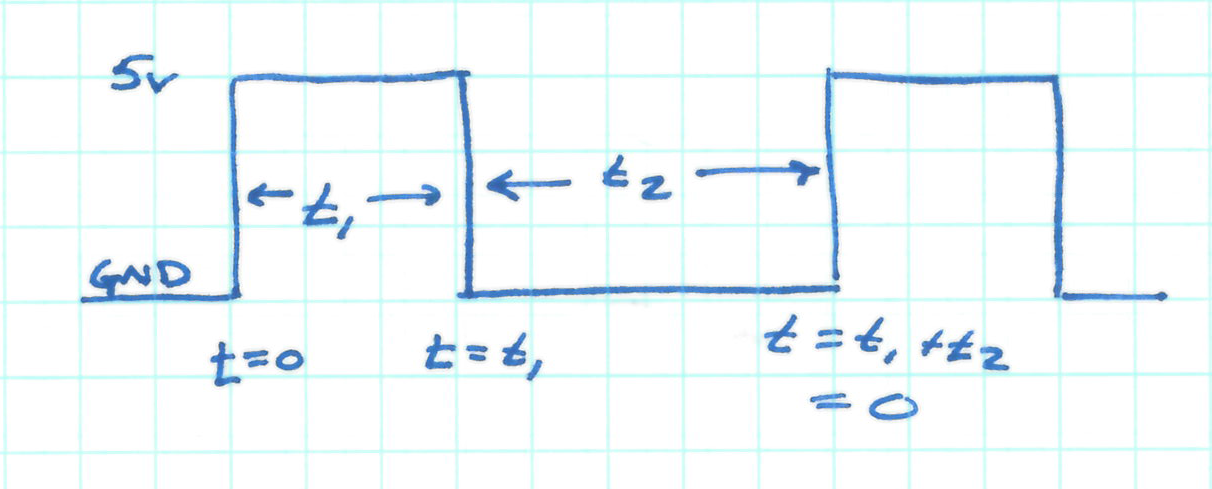}
\caption{The PWM signal is a variable length square wave.  Mathematically it can be thought of as a sum of $\Theta(t)$ functions. The microcontroller's ``rail" voltage, $V_{cc}$ is assumed to be the ouput level, ie $V_{cc}=5v$.}    
\label{fig:PWM-wave}
\end{figure}
\end{center}

If we want the LED to be a bit brighter, the on state can be held for $3ms$ while the off state remains at $1ms$.  Or, if we need the LED to be very dim, we can hold the on state for $500us$ and then the off state for $3500us$.  Formally, we can implement this as a ``Pulse Width Modulation" routine, where the voltage the microcontroller sends is on/high/$5v$ for a time $t_1$ and then off/low/$0v$ for a time $t_2$.  The overall period of the cycle is $T=t_1+t_2$, graphically shown in figure \ref{fig:PWM-wave}, or mathematically, one cycle is given,

\bea
v(t)&=& 5v~\textrm{for}~0<t<t_1 \nonumber \\
&=&0v~\textrm{for}~t_1<t<t_1+t_2.\label{eq:pwm_1}
\eea 

Note, equation \ref{eq:pwm_1} is usually periodic, as $T=t_1+t_2$ is usually quite short compared to the relaxation time $T_R$ of the phenomena we're trying to drive.  What is a relaxation time?  When power is removed, a computer cooling fan will take a few seconds to spin down to rest.  The length of time probably depends on the fan's age, accumulated dust, and the quality of the bearings in the fan.  Thus for the fan, $T_R\approx5s$.  In your eye, $T_R$ is probably on the order of a few milliseconds, because if a phenomena is faster than $\approx5ms$, the motion appears smooth, while motions slower than this time are seen discretely.  

Technically, a better definition of $T_R$ is to look for a exponential response, $\exp[-t/T_R]$, in the behavior of an object when the driving force is removed.  To be ``smoothly" controlled by PWM, the PWM period, $T=t_1+t_2$ should be significantly less than exponential relaxation/decay time, ie $T\ll T_R$.

In an automobile's internal-combustion engine, one of the pistons fires with each engine revolution.  If the car's motor is running at $2500RPM$, a piston will fire and push on the crankshaft at a rate of $\approx42$ combustions each second.  So, for an engine, $T\approx \frac{1s}{42~pushes}\approx{24ms}$.  Automotive engines are built with large, heavy, steel flywheels attached to the crankshaft.  The flywheel's angular momentum gives the system a $T_R\gg 24ms$, which is part of what leads to a smooth-running car.  

The electrical PWM function, equation \ref{eq:pwm_1}, has an average value, $\langle V \rangle$, that is straight-forward to compute.  Skipping the integral, we have: 
\be
\langle V \rangle = \frac{1}{t_1+t_2}\left(V_{cc}\cdot t_1 + 0v\cdot t_2\right) = V_{cc}\frac{t_1}{t_1+t_2}
\label{eq:raw_avg}
\ee
The amount of time the wave is held high corresponds to a ``duty-cycle" level, $d=\frac{t_1}{t_1+t_2}$, which simplifies the expression for average output, $\langle V \rangle=d \cdot V_{cc}$.

\section{A basic PWM implementation in C}
Implementation of equation \ref{eq:pwm_1} in Arduino C can take the form of a for loop with repeated delays.  The loop starts with the output pin high, and then after delays corresponding to $t_1$ have elapsed, the output pin is pulled low for the remainder of the PWM cycle.   

\lstset{language=C,numbers=left, numberstyle=\tiny, stepnumber=2, numbersep=5pt}
\begin{lstlisting}[caption={Simplest PWM routine},label=la:first]
int output_pin = 13;
void setup() {
  pinMode(output_pin, OUTPUT);
  digitalWrite(13, LOW);
}

void loop() {
  int j;

  int num_pwm_bins = 32;
  int brightness = 5; 

  digitalWrite(output_pin, HIGH);
  for (j = 0; j < num_pwm_bins; j++) {
    if (j == brightness) {
      digitalWrite(output_pin, LOW);
    }
    delayMicroseconds(20);
  }
}
\end{lstlisting}

The rough function of this code follows.  Pin 13 is used to output the PWM signal. In the Arduino environment the loop() function, line 7, runs in an infinite loop.  Each iteration of loop() generates a single PWM wave, equation \ref{eq:pwm_1}, and is broken into 32 sections via a for loop, line 14.  Each section consists of a $20us$ delay and a check to see if the duty-cycle for the pwm wave has been met.  The function generates a duty cycle of $d \approx \frac{brightness}{num\_pwm\_bins}$, because after ``brightness" number of sections have elapsed, line 15, the output pin is pulled low.  

In this implementation, $t_1\approx 20us\cdot5=100us$, $t_2\approx 20us\cdot27=540us$, and $d\approx \frac{5}{32}\cdot 5v \approx 0.78v$.  Note that this implementation's output has a granularity of $\frac{1}{32}\cdot 5v \approx 0.15v$, ie we can't specify an output level with precision finer than $0.15v$.  If we wanted to specify an average output voltage at the $0.001v$ level, the ``num\_pwm\_bins" variable, line 9, would need to satisfy the description, 
$0.001v=\frac{1}{num\_pwm\_bins} \cdot V_{cc}$.  
At $V_{cc}=5v$, this would mean $num\_pwm\_bins \approx 5000$.

Implicitly then, there's a trade-off between the granularity or precision of a PWM signal and time it will take to implement.  A signal with $0.001v$ precision takes $5000~steps\cdot\frac{20us}{step}\approx0.1s$ to implement.  $0.1$ seconds is an eternity in the time-frame of a microcontroller running at $20MHz$!

\section{Averaging PWM output with a low-pass filter}
\label{sec:RC_filter}

\begin{center}
\begin{figure}
\includegraphics[width=\columnwidth]{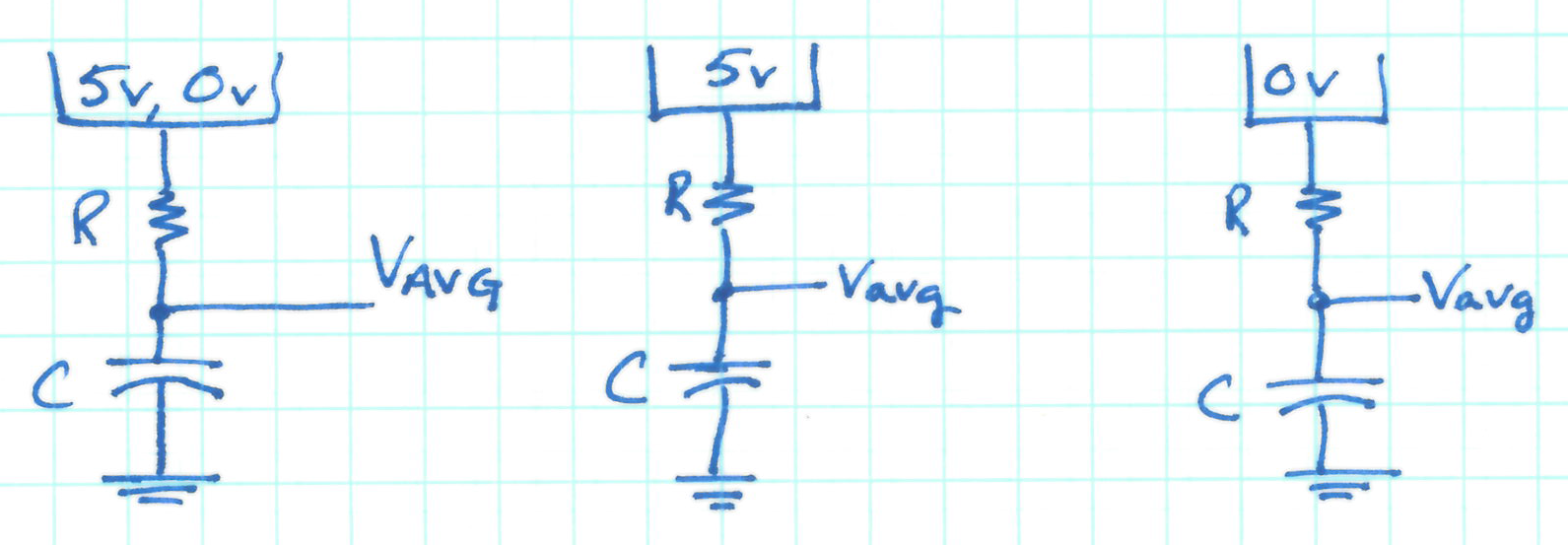}
\caption{A PWM averaging circuit, using a resistor $R$ and capacitor $C$.  In some contexts this arrangement is described as a low-pass filter.  The circuit is redrawn under charging (PWM HIGH) and discharging (PWM LOW) conditions.     }    
\label{fig:pwm_avg_circ}
\end{figure}
\end{center}

When used electrically, PWM signals are often averaged by a resistor-capacitor filter circuit, shown in figure \ref{fig:pwm_avg_circ}.  In steady-state conditions, the capacitor averages out the pulsed input by storing and releasing electric charge in the normal ``RC" fashion.

Students are normally introduced to capacitor charging and discharging in their second semester of University Physics.  See for example, OpenStax University Physics, vol 2, section 10.5 - RC Circuits.

The Kirchhoff voltage loop for the filter can be written in the following way,
\bea
V_{cc} -iR -Q/C &=& 0,~ \textrm{for}~ 0<t<t_1 \nonumber \\ 
 -iR -Q/C &=& 0,~ \textrm{for}~ t_1<t<t_1+t_2. \label{eq:kirch}
\eea
The solutions to this set of equations, $Q_1$ during $t_1$ and $Q_2$ during $t_2$, are
\bea
Q_1(t) &=& A \exp[-t/\tau]+B,~\textrm{and}\nonumber \\ 
Q_2(t) &=& D \exp[-t/\tau]. \label{eq:sln}
\eea
The current in the resistor is the time-derivative of the capacitor charge, $i=\frac{dQ}{dt}$, and plugging the solutions, equation \ref{eq:kirch}, into the voltage equations, \ref{eq:sln}, give the relations,
\bea
\tau&=&RC \nonumber \\
B &=& C~V_{cc}. \label{eq:params}
\eea
At this point, amplitudes $A$ and $D$ remain undefined.  The amount of charge on the capacitor must be continuous, so two additional equations of constraint are:
\bea
Q_1(0)&=&Q_2(t_1+t_2),~\textrm{and,} \nonumber \\
Q_1(t_1)&=&Q_2(t_1). \label{eq:boundary_cond}
\eea
When these boundary conditions are applied to the equation \ref{eq:sln}, you can determine $A$ and $D$ to be:
\bea
A &=& V_{cc}\cdot C \left( \frac{\exp[-t_2/\tau]-1}{1-\exp[-(t_1+t_2)/\tau]}\right),~\textrm{and,}\nonumber\\
D &=& V_{cc}\cdot C \left( \frac{\exp[t_1/\tau]-1}{1-\exp[-(t_1+t_2)/\tau]} \right)
\eea 

\begin{center}
\begin{figure}
\includegraphics[width=\columnwidth]{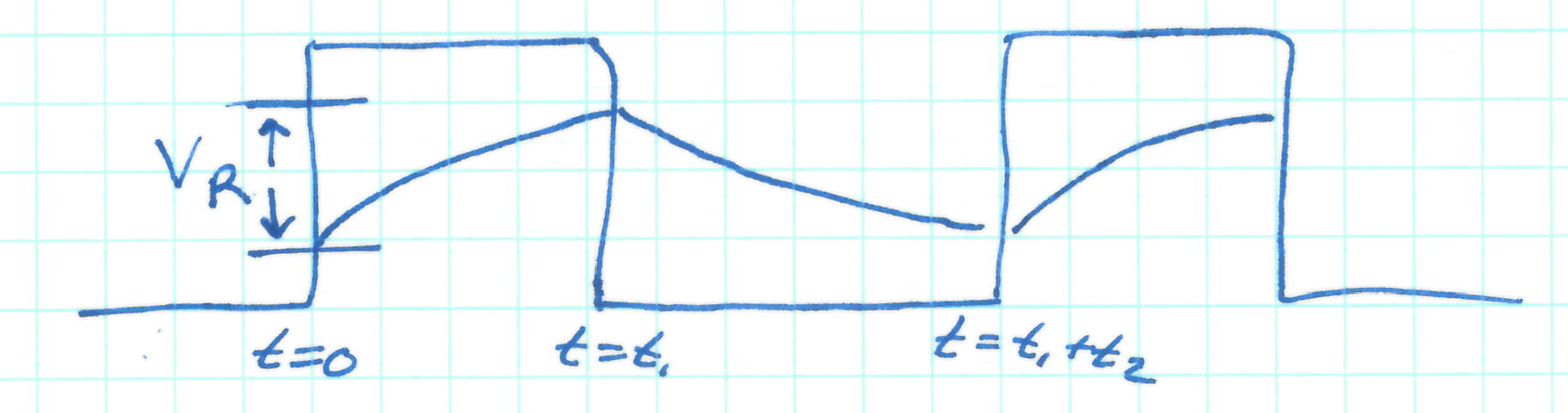}
\caption{The typical waveform that is seen at the $V_{avg}$ low-pass filter output from figure \ref{fig:pwm_avg_circ}.  This ``shark-fin" wave comes from a periodic charge-discharge cycle that's applied by the PWM output to the RC series circuit.}    
\label{fig:PWM-avg_wave}
\end{figure}
\end{center}

\section{Check Your Understanding Questions}
\begin{enumerate}
\item For each of the following systems identify and estimate the (PWM) push and the averaging system.  Additionally estimate PWM period, $T$, and system relaxation time, $T_R$.
\begin{enumerate}
\item The two cycle engine in a chainsaw or outboard motor.
\item A muscle fiber in your bicep.
\item The human digestion system.
\item A variable speed motor in a hot air furnace blower.
\item A dimmable 120VAC light bulb.
\item Constituents calling a congressperson's office. 
\item A bicycle.
\item Human locomotion, ie swimming or walking.
\end{enumerate}
\item If the voltage across the LED in figure \ref{fig:PWM-POT} is $2.2v$, what fraction of the system's power actually makes it to the LED to be used 
to generate light?
\cite{UMN_MXP}
\end{enumerate}

\section{Lab 1}
These problems are written in an ascending order of difficulty.  Not all need to be done.  This took about 4 hours of in-class lab work.
\begin{enumerate}

\item How closely does the the PWM listing given, \ref{la:first}, match the timing specifications in \ref{eq:pwm_1}?  Use an oscilloscope to inspect the output signal. Deviations from the timing model probably come from latency in the for loop.  

\item How can you rewrite code listing \ref{la:first} to better match the PWM timing specifications in equation \ref{eq:pwm_1}? Hint, call delay only twice per PWM cycle.  \textbf{Solution:}
\lstset{language=C,numbers=left, numberstyle=\tiny, stepnumber=2, numbersep=5pt}
\begin{lstlisting}[caption={PWM routine with better timing},label=la:second]
int output_pin = 13;
void setup() {
  pinMode(output_pin, OUTPUT);
  digitalWrite(13, LOW);
}

void loop() {

  int num_pwm_bins = 32;
  int brightness = 10; 
  int cycle_length = 20; 
  int t1 = cycle_length * brightness;
  int t2 = cycle_length * 
    (num_pwm_bins - brightness);

  while (1) {
    digitalWrite(output_pin, HIGH);
    delayMicroseconds(t1);
    digitalWrite(output_pin, LOW);
    delayMicroseconds(t2);
  }
}
\end{lstlisting}
It is informative to compare these two PWM functions in an oscilloscope.  The latency implicit in the for loop and repeated delay calls in listing \ref{la:first} becomes quite obvious.

\item 
Write a program that generates a PWM signal that has a minimum delay of $10us$ ($0.1 MHz$), and a period of $10ms$.  Implicitly, this is a duty level that can be specified to $1$ part in $1000$.  Document the accuracy of your routine with a multimeter and an oscilloscope.  
Specifically, what is the finest resolution of average voltage you can specify between adjacent duty levels?  
Also, what are the on-off times in the oscilloscope output?  How closely do they match the millis/micros delay times in your code?

\item 
Fold your PWM routine into a function.  It should take some standard outputs that you define, eg, carrier period, duty level, duration of pulse, etc.  Make sure the function works - it is a good way to clarify your thinking and code!

\item Use a capacitor and a load resistor, per figure \ref{fig:PWM_meas}, connected between the PWM lead and ground, to average out the PWM signal.  Inspect the averaged and unaveraged signal on the oscilloscope.  Is the capacitor’s effect uniform across duty levels?  This works best if the capacitor is large.

\begin{center}
\begin{figure}[h]
\includegraphics[width=\columnwidth]{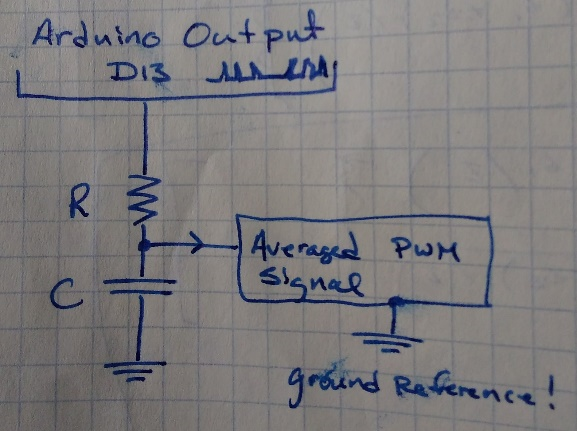}
\caption{This is the typical wiring diagram we've used in lab.  An Oscilloscope is used to measure te averaged PWM signal between the resistor and capacitor.  In production, for example, when playing music, an operational amplifier (opamp) can be used to amplify the signal produced at the Resistor-capacitor junction.}    
\label{fig:PWM_meas}
\end{figure}
\end{center}

\item Use your PWM routine to create a sawtooth function, with $T = 2sec$.  You'll have to call your PWM function repeatedly with an intentionally chosen increase each time the PWM function is run.  Inspect the function with an oscilloscope with and without a capacitor to average out the signal.  How accurately have you created a ``perfect” sawtooth?  Note, in this problem you are basically creating an arbitrary function generator.

\item
Use your PWM routine and a 3904 transistor (or similar) to drive an external load with your sawtooth.  (A DC computer cpu fan is easy, but there are other options, eg a speaker at $20kHz$ \ldots).  Note, if you want to drive a fan, the Blum Arduino book has a reasonable recipe on pp 65-69, \cite{Blum}.  One way to check your work is to listen to the fan as the sawtooth winds up and down.  
\end{enumerate}

\section{Lab 2}
This took about 4 more hours of in-class lab time (and the students worked on this outside of class as well!)

\begin{enumerate}
\item 
Generate a PWM wave with average value of $3.27\pm0.01$ volts, measured with a multimeter.  Reminder, the PWM function we’ve been working with is HIGH ($5.0v$) for $t1$ and LOW ($0v$) for $t2$. The cycle is periodic at a time $t=t1+t2$.  How many intervals do you need to break the PWM cycle into to achieve $\pm0.01$ volt resolution?
 
\item 
Generate an periodic oscillating function that is $1.22v$ for $100ms$ and $3.77v$ for $200ms$ (both $\pm0.01v$). Note, for this you'll want to use a resistor-capacitor filter like that shown in figure \ref{fig:PWM_meas}.

\item 
Starting with Kirchhoff loops for the filter when the system in the OK and OFF states, work out the mathematical details of \ref{sec:RC_filter}.  You'll need to set up two differential equations, propose and try out two solutions, and apply boundary conditions at $t=0$, $t=t1$, and $t=t1+t2$ to solve for all undetermined variables in the solution.

\item
Build the circuit in figure \ref{fig:PWM_meas} with known $R$, $C$, $t1$, and $t2$ values and capture the filtered output via an oscilloscope for a duration of $\approx2\left(t1+t2\right)$.  Compare the oscilloscope output to a plot (in Mathematica or similar) for $Q_1$ and $Q_2$ above (recall, $Q=CV$ for capacitors).  There should be good agreement!

\item 
The ``Ripple Voltage," $V_R$, in this PWM scheme is the total variation in output voltage in the averaged PWM signal.  $V_R$ is shown graphically in figure \ref{fig:PWM-avg_wave}.  Using $Q_1$ and $Q_2$, equation \ref{eq:sln} above, create an algebraic expression for this ripple voltage, $V_R$.  Then, using several different $t1$, $t2$, $R$, and $C$ values, test the accuracy of your prediction for $V_R$. 

\item 
Application Question: if you hold $R$ and $C$ constant, are there $t1$ or $t2$ values that make the ripple voltage larger?  (Application, are the PWM levels at which this averaging routing makes a noisy, unreliable signal? Or, if $t1+t2$ is fixed, are there $t1$ or $t2$ values that make the averaging element produce a clean, smooth signal?)

\item 
Application Question: If you want the ripple voltage to be no more than $10\%$ of the ideal output, $V_{CC}\frac{t1}{t1+t2}$, what constraint should be imposed on $R$ and $C$?  Back up your answer with data from lab.

\item 
Use the averaged PWM output to create a function that will play a (musical) note of frequency $f$.  A musical note can be a sine wave, $V\left(t\right)=Sin\left[2\pi f t\right]$.  The pitch (high-ness) of the note is dictated by its frequency, $f$, where $f=1/T$, and T is the period, or ``length in time” of one cycle of the sine wave. Like the sawtooth before, you'll need to make repeated calls to the PWM function to approximate the shape of a sine curve.  Note as well that your sinewave will be generated with a DC off-set,  which you could remove with an opamp.   A SEEED Grove buzzer, \cite{buzzer}, is an easy way to play this note.  Here's a list of Piano key frequencies, \href{https://en.wikipedia.org/wiki/Piano_key_frequencies}{https://en.wikipedia.org/wiki/Piano\_key\_frequencies}

\item Extra credit:  Use your musical note function to play a song on a speaker. We have low power buzzers in the SEEED Grove kits, or you can use a speaker, but you might need to use an op-amp (LF411) to scale the output.  Happy Birthday or Edelweiss are nice choices.     

\end{enumerate}
 

\end{document}